\documentclass[11pt]{article}

\usepackage{amsmath,amssymb,bbm}
\usepackage{graphicx}
\usepackage{a4wide}

\newcommand{\sst}{\scriptscriptstyle}
\newcommand{\Ep}{E^{\sst \|}}
\newcommand{\Eo}{E^\bot}

\newcommand{\p}{{\sst \|}}
\newcommand{\e}{\text{\bf e}}

\title{Quasiperiodic Tilings:\\ A Generalized Grid--Projection Method}

\author{
  V.\ E.\ Korepin\\[1mm] 
  {\it\small Leningrad Department of the Steklov Mathematical Institute,}\\
       {\it\small Fontanka 27, 191011 Leningrad, USSR}
\and
  F.\ G\"ahler\footnote{Present address: {\it D\'epartement de Physique 
    Th\'eorique, Universit\'e de Gen\`eve,
    24 Quai Ernest Ansermet, CH-1211 Gen\`eve 4, Switzerland}}
  \ and J.\ Rhyner\footnote{Present address: 
    {\it Asea Brown Boveri, Corporate Research, 
    CH-5405 Baden, Switzerland}}\\[1mm]
  {\it\small Institute for Theoretical Physics, ETH H\"onggerberg,
       CH--8093 Z\"urich, Switzerland}
}

\date{}

\begin{document}

\maketitle

\begin{abstract}
We generalize the grid--projection method for the construction of
quasi\-periodic tilings. A rather {\it general} fundamental domain of
the associated higher dimensional lattice is used for the construction
of the acceptance region. The arbitrariness of the fundamental domain
allows for a choice which obeys all the symmetries of the lattice,
which is important for the construction of tilings with a given
non-trivial point group symmetry in Fourier space. As an illustration,
the construction of a 2d quasiperiodic tiling with twelvefold
orientational symmetry is described.
\end{abstract}

\setcounter{section}{-1}

\section{Introduction}

The interest for non-periodic tilings first arose from problems in
mathematical logics (Wang 1965, Robinson 1971).
However, since Penrose's invention of his well
known non-periodic tilings (Penrose 1974, 1979, Gardner 1977),
the motivation has changed to the study of geometrical questions related
to such tilings. J.~Conway (see M.~Gardner 1977) and N.~G.~de~Bruijn (1981)
have played a dominant r\^ole in this field.

We define a tiling as a covering of the plane by translations
of a finite number of polygons with no holes or overlaps.
Note however that, depending on the context, other definitions may be more
appropriate, see e.~g.\ Gr\"unbaum \& Shepard (1987).
Here we are interested in quasiperiodic tilings. By this we mean:
\begin{itemize}
\item[(i)]The tiling is not periodic: There exist no translations
(exept the identity) which leave the tiling unchanged.
\item[(ii)]If we put a $\delta$-function to each vertex of the tiling,
the Fourier transform of the resulting structure is a sum of
$\delta$-peaks, whose positions are integer linear combinations of a
finite set of vectors $\{\text{\bf k}_1,\ldots,\text{\bf k}_n\}$:
\begin{equation}
F(\text{\bf k})=\sum_{\ell \in \text{\bf Z}^n}w_{\ell_1,\ldots,\ell_n}
\delta (\text{\bf k}-\sum_{i=1}^n\ell_i\text{\bf k}_i).\tag{0.1}
\end{equation}
\item[(iii)]Any finite part of the tiling appears infinitely often in the
tiling.
\end{itemize}

\noindent
Condition (iii) is often dropped, but the tilings we will consider
have this property.

Quasiperiodic tilings may have symmetries in Fourier space which are
incompatible with a periodic structure and do therefore not occur for crystals.
A famous example is the Penrose tiling with fivefold symmetric Fourier
transform. This has led physicist's interest to quasiperiodic tilings, since
icosahedrally symmetric diffraction patterns of an $Al$-$Mn$ alloy, observed
by Shechtman et al.\ (1984), could be explained in terms of a threedimensional
version of the Penrose tiling (Mackay 1981, Duneau \& Katz 1985, Elser 1986,
Kalugin, Kitaev \& Levitov 1986, Levine \& Steinhardt 1984).
More information on the symmetry of quasiperiodic tilings and the
connection to physics can be found in the Les Houches proceedings (1986).

The main concern of this paper will be a generalization of the grid-projection
method used by various authors (Duneau \& Katz 1985, Elser 1986,
G\"ahler \& Rhyner 1986, Kalugin, Kitaev \& Levitov 1986, Korepin 1986,
Kramer \& Neri 1984, Socolar, Steinhardt \& Levine 1985). Our
algorithm projects part of a $n$-dimensional lattice $\Gamma$ in $E^n$ onto
an irrationally embedded $d$-dimensional subspace. It is based on a
periodic tiling of $E^n$ by copies of a rather
{\it general} fundamental domain of $\Gamma$,
as opposed to G\"ahler \& Rhyner (1986), who considered only fundamental
{\it parallelotopes}. This extension allows for the choice of a fundamental
domain whose closure obeys all the symmetries of the lattice,
which is important for the
construction of tilings with specified symmetry in Fourier space.
The tilings so obtained are then no more tilings by parallelotopes
alone, but can contain any kind of convex polytopes. These
tilings will not be the final step however. We rather prefer to consider
their Voronoi partitionings, since these seem to have more relevance to physics
(Jaric 1986, Henley 1986).

The outline of this paper is as follows. After briefly reviewing the
concepts of Voronoi partitioning and tilings by fundamental domains in
sections 1 and 2, we present our generalized algorithm in section~3.
In section~4 we prove that the tilings constructed in section~3 are
indeed quasiperiodic in the sense explained above. Finally, section~5 is
devoted to an example to illustrate these techniques.

\section{Tilings by parallelotopes and Voronoi domains}

The standard grid-projection method (G\"ahler \& Rhyner 1986) yields
tilings of the Euclidean space $E^d$
by a finite set of parallelotopes. Possible sets of paralellotopes can be
obtained as follows: Let $\{\text{\bf t}_i\}_{i=1,\ldots,n}$ be a set of $n$
vectors in $E^d$ ($n>d$). Any subset of $d$ linearly independent vectors
of this set spans a parallelotope (see next section). With the parallelotopes
obtained in this way $E^d$ can be tiled both periodically and quasiperiodically.

It should be noted that the set of vertices of such a tiling is a special
case of a {\it Delauney (r,R)-system} (Delauney 1937),
which is a point set $\{v_j\}$ in $E^d$ with the following two properties:
\begin{itemize}
\item[(i)]The minimal distance between any two points of the system is $r>0$.
\item[(ii)]Inside or on the surface of any ball of radius $R$, no matter
where we put its center, there is at least one point of the system.
\end{itemize}

\noindent
With each (r,R)-system we associate its {\it Voronoi partitioning} of $E^d$,
which divides $E^d$ into {\it Voronoi domains}, also called Dirichlet or
Wigner-Seitz cells. The Voronoi domain associated with $v_j$
is a convex polytope and consists
of all points of $E^d$ whose distance to $v_j$ is not larger than the distance
to any other point $v_i\neq v_j$ in the system. Note that according to this
definition a Voronoi domain is a closed set, which means that boundary
points belong to two or more Voronoi domains.
An (r,R)-system has in general infinitely many different
Voronoi domains, but a quasicrystal can have only finitely many (up to
shifts), see section~3. A (periodic) lattice has even only one type of Voronoi
domain.

The Voronoi domain
around a point $v_j$ of an (r,R)-system in $E^d$ can be constructed as
follows. Consider the set of all vertices $v_i$ inside a closed ball of
radius $2R$ centered at $v_j$. For each point $v_i$ in this set, construct
the $(d-1)$--dimensional hyperplane perpendicular to the segment $v_j-v_i$
and passing through its midpoint $\frac12 v_i + \frac12 v_j$. Each of these
hyperplanes cuts
$E^d$ into two halfspaces. The Voronoi domain $v_j$ is the intersection of all
those halfspaces which contain $v_j$.

\section{Fundamental domains of a lattice}

Consider a lattice $\Gamma$ in $E^n$ generated by $n$ linearly independent
vectors $\e_1,\ldots,\e_n$. A {\it fundamental parallelotope} $F_p$ of
$\Gamma$ is the set
\begin{equation}
F_p=\sum_{i=1}^n \lambda_i\e_i,\quad 0\le\lambda_i<1. \tag{2.1}
\end{equation}
Since a lattice has infinitely many lattice basis, it has also infinitely
many fundamental parallelotopes. A fundamental parallelotope is a special
case of a {\it fundamental domain} $F$ of a lattice, which is a measurable
set with the following two properties:
\begin{itemize}
\item[(i)]The translates of $F$ by all lattice vectors of $\Gamma$
cover $E^n$ with multiplicity one.
\item[(ii)]$F$ contains exactly one point of the lattice.
\end{itemize}

\noindent
It should be noted that a fundamental domain is neither closed nor open,
only a part of the boundary belongs to $F$. In the following we will
restrict ourselves to fundamental domains whose closure is a convex
polytope. Particularly interesting is a fundamental domain whose
closure is the Voronoi domain. From property (ii) it follows that
with each fundamental domain $F$ a unique lattice point
$\gamma(F)$ is associated. This will become important later.
                                              
\section{The generalized grid-projection method}

Let us decompose the space $E^n$ containing the lattice $\Gamma$ into
two orthogonal subspaces, $E^n=\Ep\oplus\Eo$.
We will assume in the following that this decomposition is irrational, i.\ e.\
neither $\Ep$ nor $\Eo$ contain any lattice vectors of $\Gamma$.
Let $F_0$ be a fundamental domain of $\Gamma$, and $\mathcal{P}_\Gamma (F_0)$ the 
partitioning of $E^n$ into all $\Gamma$-translates of $F_0$. The closures 
of the $\Gamma$-translates of $F_0$ will be called the cells of the 
partitioning. We assume that the cells are convex polytopes, and that 
the partitioning $\mathcal{P}_\Gamma (F_0)$ is face-to-face, i.\ e.\ if 
two cells have a non-zero intersection, then this intersection is a 
(common) face of these two cells. Note that the 
Voronoi partitioning is always face-to-face. If a cell is not equal 
to a Voronoi domain of the lattice, we moreover assume that the 
partitoning is generic in the sense that a face of dimension $m$ is
contained in exactly $n-m+1$ cells. The dual of such a partitioning is simplicial,
i.e. all cells (and their faces) are simplices. Next,
consider an $d$-dimensional (affine) subspace $E$ of $E^n$ which is
parallel to $\Ep$. We assume that $E$ is located at a generic position, so that
only faces of dimensions $n$-$d$ to $n$-$1$ of the cells have non-zero 
intersection with $E$.

The generalized projection method now is described as follows.
The vertices of the tiling are obtained by projecting orthogonally onto
$E$ the set $W$ of lattice points whose associated fundamental
domain has a non-zero intersection with $E$:
\begin{equation}
W=\{\gamma(F)\,|\,F\cap E \not = \emptyset,
  F\in \mathcal{P}_\Gamma(F_0)\}. \tag{3.1}
\end{equation}
Next, we have to divide $E$ into tiles by specifying all their faces of 
dimensions up to $d-1$. The 1-dimensional ``faces'' are obtained by
connecting all those vertices by a straight line whose associated cells
share a common face of dimension $n-1$ which cuts $E$. If $d=2$, the 
tiling is then completely specified. For $d>2$ however, the situation is
somewhat more complicated. 
Those lattice points whose associated cells share a face of
dimension $k$ are the corners of a convex polytope of dimension $n-k$.
For a generic partitioning this is evident, for there are always exactly
$n-k+1$ such points. For the Voronoi partitioning, we can argue differently.
The points whose cells share the $k$-face under consideration can all be
connected by a chain or net of straight lines each of which is perpendicular
to an $(n-1)$-face containing the $k$-face and thus perpendicular to the 
$k$-face itself. Therefore, all these points are contained in a single
plane of dimension $n-k$ perpendicular to the $k$-face. Hence, in both
the Voronoi and the generic case we can build the $(n-k)$-dimensional
polytope dual to a given $k$-face. If now the $k$-face cuts $E$, we project
its dual polytope to $E$. In this way we obtain a prescription for the
subdivision of $E$ into tiles. Note that with each projected dual of a 
$k$-face also all its boundaries are projected, since the $k$-face is
contained in the corresponding $(k+1)$-faces which cut $E$ too.

The same tiling can also be obtained as the dual of a grid $G$. This grid
is given by the intersection of the union of the boundaries of all
cells of the partitioning with the subspace $E$. The grid divides
$E$ into convex polyhedral cells, called {\it meshes}, the faces of which are
the intersections of $E$ with the $(n-1)$-dimensional faces of the cells 
of the partitioning. Each mesh of the grid corresponds to a cell which cuts 
$E$. Therefore, with each mesh we can
associate the projection of the corresponding lattice point, and two
lattice points belonging to meshes with a common $(d-1)$-face have to be connected
by the projection of the corresponding lattice vector connecting the two
lattice points. The vertices associated with the meshes sharing a common $k$-face 
will become the corners of a $(d-k)$-face of a tile (these vertices are indeed 
contained in a $(d-k)$-plane as explained in the previous paragraph). In this 
way we see that the tiling obtained previously by projection can be reconstructed 
from the grid. According to this construction, it is the dual graph of the grid.

What is not immediately clear is whether there will be overlapping tiles,
i.\ e.\ whether the tiling is folded. Whether there are additional conditions
required to avoid overlapping, and what these conditions would be, we leave 
as an open problem. For the Voronoi case however we have some (numerical)
evidence that overlapping does not occur, and we conjecture that this
is generally true for the Voronoi case. For the classical grid method,
the necessary and sufficient non-overlapping conditions have been determined
(G\"ahler \& Rhyner 1986, de Bruijn 1986).

From the grid picture and from the periodicity of $\Gamma$ it follows
that the tiling consists only of a finite number of different tiles (up
to translation), for there are only finitely many inequivalent $(n-d)$-faces
of the cells of the partitioning which can cut $E$ (note that the type of such
an $(n-d)$-face determines which vertices belong to the associated cell).
By a similar reasoning one finds that there are only finitely many
arrangements of cells which share a common vertex, so that the Voronoi
partitioning of the tiling, as constructed in section~1, consists of a
finite number of different cells too. Bounds on the number of different
patches of radius $R$ of such a quasiperiodic tiling have been obtained
by G\"ahler (1986). This number is finite and can grow only with a fixed power
of $R$.

Let us compare our construction briefly with the algorithm proposed by
G\"ahler \& Rhyner (1986).
They consider only special fundamental domains, namely parallelotopes.
This has the disadvantage that for non-orthogonal
lattices it is impossible to choose a parallelotope which is invariant
under the whole point group of the lattice $\Gamma$. The choice of a symmetric
fundamental domain is essential for the construction of quasiperiodic
tilings which have the corresponding symmetry in Fourier space.
By allowing a more general fundamental domain, e.\ g.\ the Voronoi domain,
this deficiency is removed. This additional freedom is the main difference
as compared to G\"ahler \& Rhyner (1986). Using different ``grid-'' and ``tiling-spaces''
or including a subsequent linear transformation applied to the tiling
could of course also be incorporated into the present algorithm.

\section{Proof of quasiperiodicity}

In this section we demonstrate that the tilings constructed in the last
section satisfy the three conditions for quasiperiodicity formulated
in the introduction. Since the proof of condition (ii) is a
standard one (see e.\ g.\ G\"ahler \& Rhyner 1986, Zia \& Dallas 1985),
we restrict ourselves to conditions (i) and (iii).

First we prove non-periodicity. Let us define the projectors $P^{\sst \|}$
and $P^\bot$ projecting orthogonally onto $\Ep$ and $\Eo$ respectively.
Further, define the {\it strip S} as
\begin{equation}
S=\{\text{\bf m}+\text{\bf e}\,|\,\text{\bf m}\in M, 
  \text{\bf e}\in\Ep\},\tag{4.1}
\end{equation}
where the {\it acceptance region} $M$ is the projection $P^\bot F$
onto $\Eo$, with $F$ a translate of $F_0$ centered at $E$.
Then, we can write the set $W$ defined in (3.1) as $W=\Gamma\cap S$.
Clearly, the projection of $W$ onto $\Ep$,
$W^{\sst \|}=P^{\sst \|}W$, is the set of vertices of the tiling.
Due to the irrationality of the embedding of $\Ep$ and $\Eo$, the sets
$P^{\sst \|}\Gamma$ and $P^\bot\Gamma$ are dense in $\Ep$ and $\Eo$, and there
is a one-to-one correspondence between $\Gamma$, $P^{\sst \|}\Gamma$ and
$P^\bot\Gamma$, as well as between $W$, $W^{\sst \|}$ and $W^\bot=P^\bot W$.
Suppose that $W^\p$ is periodic, i.~e.\ $W^\p$ is invariant under a
translation $\gamma_\p$. Then $\gamma_\p$ maps vertices to vertices and is
therefore the projection of a lattice vector $\gamma$. Hence, $W^\p$ is
as well the projection of the set $W+\gamma$. Since there is a one-to-one
correspondence between $W$ and $W^\p$ this means that $W=W+\gamma$. Let us
project this equation to $\Eo$: $W^\bot=W^\bot+P^\bot\gamma$.
Since the closure of $W^\bot$ is compact, this means that $P^\bot\gamma=0$ or
$\gamma\in\Ep$, which contradicts our assumption of an irrational embedding
of $\Ep$. Therefore the tiling is non-periodic.

Now we show that every finite part $W_f^{\sst \|}\subset W^{\sst \|}$ has
infinitely many copies in $W^{\sst \|}$. Denote by $W_f$ the unique subset
of $W$ such that $W_f^{\sst \|}=P^{\sst \|}W_f$, and by $W_f^\bot$
its projection to $\Eo$. Since $E$ is at a generic position, no lattice
points are projected onto the boundary of $M$, and so $W_f^\bot$ is in the
interior
of $M$. Let $\Delta$ be the distance of $W_f^\bot$ to the boundary of $M$,
\begin{equation}
\Delta=\min_{\substack{{\text{\bf x}\in W_f^\bot}\\ {\text{\bf y}\in\partial M}}}
(|\text{\bf x}-\text{\bf y}|).\tag{4.2}
\end{equation}
For every lattice vector $\gamma$ whose projection onto $\Eo$ is inside an
open ball of radius $\Delta$, we have that the finite set
$\tilde W_f=W_f+\gamma\subset\Gamma$ projects into $M$,
$P^\bot\tilde W_f\subset M$, and therefore belongs to the strip $S$.
This means that a translation by $P^{\sst \|}\gamma$ maps
$W_f^{\sst \|}$ onto an equivalent set. Since $P^\bot\Gamma$ is dense in $\Eo$,
there are infinitely many such lattice vectors, and so the proof is completed.

\section{Example: a dodecagonal tiling}

As an application, we discuss the construction of a class of twodimensional
tilings with twelvefold symmetric Fourier spectrum. These tilings have first
been constructed by Stampfli (1986) by means of a grid. They might be relevant for
the description of quasicrystalline $Ni$-$Cr$ (Ishimasa, Nissen \& Fukano
1985, G\"ahler 1987). More details about these
and related tilings can be found in G\"ahler (1987).

The relevant lattice for our case is the diisohexagonal orthogonal primitive
lattice (Brown et al.\ 1978) in four dimensions, denoted by $\Gamma$. This
lattice has a point
symmetry group which contains the subgroup $D_{24}$. The latter is the relevant
symmetry group for our purposes. The lattice $\Gamma$ is easily constructed as
follows. Let us decompose $E^4$ into two orthogonal subspaces,
$E^4=\Ep\oplus\Eo$. $\Ep$ is the space onto which we will project. Let
$\{\text{\bf e}_1,\ldots,\text{\bf e}_{12}\}$ be a star of twelve vectors
in $E^4$ such that their projections onto $\Ep$ and $\Eo$ are given by
\begin{equation}
\aligned
\e_i^{\sst \|}&=(\cos(\pi(i-1)/6), \sin(\pi(i-1)/6))\\
\e_i^\bot&=(\cos(5\pi(i-1)/6), \sin(5\pi(i-1)/6))
\endaligned
\tag{5.1}
\end{equation}
with respect to Euclidean coordinates in $\Ep$ and $\Eo$. The vectors
$\{\e_5,\ldots,\e_{12}\}$ can be expressed as integer linear combinations
of the remaining four vectors, and since $\{\e_1,\ldots,\e_4\}$ are
rationally independent, the set $\{\text{\bf e}_1,\ldots,\text{\bf e}_{12}\}$
generates a 4-dimensional lattice which will be identified with $\Gamma$.

In this example, we choose a fundamental domain whose closure is the Voronoi
domain. Therefore we have to construct the Voronoi partitioning of $\Gamma$.
We note that the space spanned by $\e_1$ and $\e_3$, denoted by $E^a$,
is orthogonal to the space $E^b$ spanned by $\e_2$ and $\e_4$. The vectors
$\e_1$ and $\e_3$ generate a 2d regular hexagonal lattice $\Gamma^a$ in $E^a$,
and $\e_2$ and $\e_4$ generate a corresponding lattice $\Gamma^b$ in $E^b$.
Therefore, $\Gamma$ is given as an orthogonal sum of two 2d regular hexagonal
lattices,
\begin{equation}
\Gamma=\Gamma^a\oplus \Gamma^b.\tag{5.2}
\end{equation}
Next we recall the fact that in such a case the Voronoi domain of $\Gamma$ is
given by the topological product of the two Voronoi domains of $\Gamma^a$
and $\Gamma^b$,
\begin{equation}
V=V^a\times V^b,\tag{5.3}
\end{equation}
which of course are regular hexagons. Let $H^a$ and $H^b$
be the hexagon nets given by the boundaries of all Voronoi domains of
$\Gamma^a$
and $\Gamma^b$ respectively. Then the union of the boundaries of all Voronoi
domains of $\Gamma$ is given by
\begin{equation}
N=(H^a\times E^b)\cup(E^a\times H^b),\tag{5.4}
\end{equation}
i.~e.\ $N$ is the union of two orthogonal arrays of hexagonal ``tubes''.
Let $E$ be a generic plane parallel to $\Ep$. The {\it grid}, i.\ e.\ the
intersection of $N$ with $E$, is then the union of the intersections of the
two arrays of tubes, which are both regular hexagon nets, turned with respect
to each other by $30^\circ$ (see Fig.~1). The elementary hexagons of these
nets are by a factor of two
larger than the projections of the Voronoi domains of the hexagonal lattices.
The relative positions of the two nets are determined by the position of $E$.

For the construction of Stampfli's tilings, the two algorithms discussed in
section~3 now read as follows:
\begin{itemize}
\item[A:]{\it Projection construction}\newline
Project the center of all those Voronoi domains of $\Gamma$ onto $\Ep$
which cut $E$.
Connect all those points by a straight line whose Voronoi domains
have a face in common which cuts $E$.
\item[B:]{\it Grid construction}\newline
With each mesh of the grid $N\cap E$, associate a vertex of the tiling.
If the meshes of two vertices have a common face, these vertices are
connected by a line of unit length which is perpendicular to this face.
This is Stampfli's presciption.
\end{itemize}

\begin{figure}
\centerline{
  \hbox to 0.49\textwidth{
    \includegraphics[width=7.2cm]{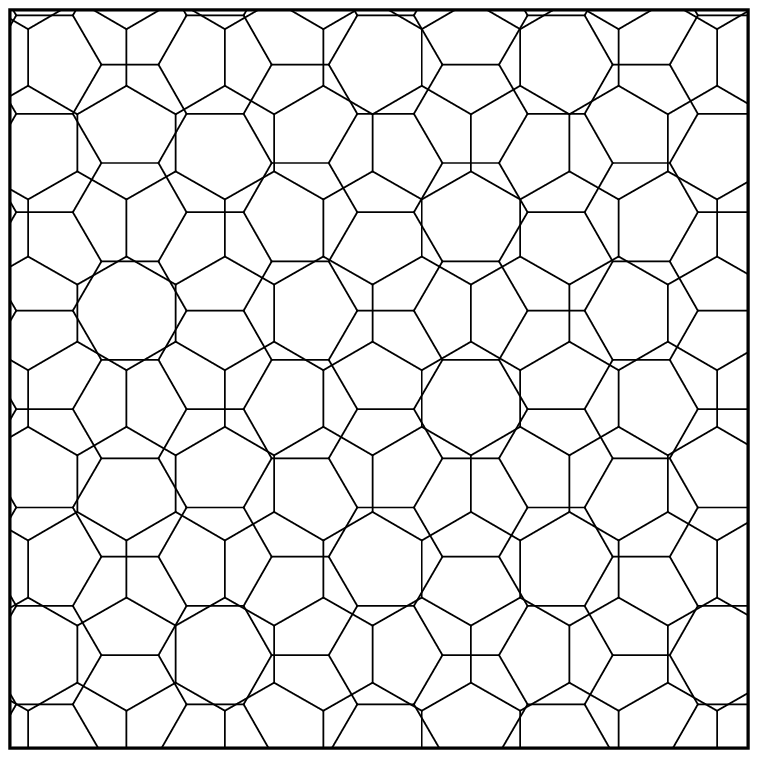}
  }  
  \hbox to 0.49\textwidth{
    \includegraphics[width=7.2cm]{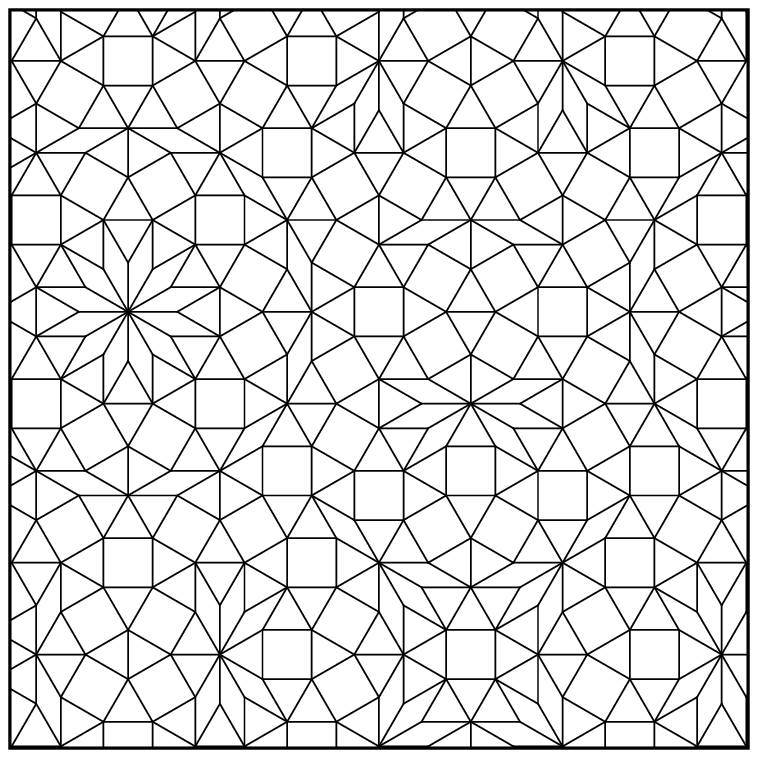}
  }
}  
\medskip
\centerline{
  \hbox to 0.49\textwidth{
    \hfil{\small Fig.~1: Grid given by two hexagon nets}\hfil
  }
  \hbox to 0.49\textwidth{
    \hfil{\small Fig. 2: Quasiperiodic tiling dual to grid of Fig.\ 1}
  }
}
\end{figure}

\noindent
A tiling constructed in this way is shown in Fig.~2.

Let us finally note that this is a particularly simple example, due to the
fact that $\Gamma$ is an orthogonal sum of two 2d lattices. From this
it follows
that the grid $N\cap E$ is the union of two simple periodic grids. In the
general case, $N\cap E$ would be very complicated, but nevertheless
all our constructions would go through as well.
\bigskip

\noindent
{\bf Acknowledgements:} J.\ R.\ is deeply indebted to the
L.~D.~Landau Institute in Moscow for the warm hospitality during the time
when part of this work has been done, and to the USSR Academy of Sciences for
financial support. F.~G.\ would like to thank Peter Stampfli for valuable
discussions concerning the example in section~5.

\section*{References}

\noindent
Brown, H., B\"ulow, R., Neub\"user, J., Wondratschek, H.\ \&
Zassenhaus, H.\ (1978). {\it Crystallographic Groups of Four-Dimensional
Space}, Wiley-Intersci\-ence (New York).

\noindent
de Bruijn, N.\ G.\ (1981). {\it Proc.\ Ned.\ Akad.\ Wetensch.\ Ser.\ A }
{\bf 43}, 39-66.

\noindent
de Bruijn, N.\ G.\ (1986). {\it J.\ Physique, }{\bf 47}, C3-9--C3-18.

\noindent
Delauney, B.\ N.\ (1937). {\it Usp.\ Mat.\ Nauk.\ } {\bf 3}, 16-62.

\noindent
Duneau, M.\, \& Katz, A.\ (1986). {\it Phys.\ Rev.\ Let.\ }{\bf 54},
2688-2691.

\noindent
Elser, V.\ (1986). {\it Acta Cryst.\ A}{\bf 42}, 36-43.

\noindent
G\"ahler, F.\ \& Rhyner, J.\ (1986). {\it J.\ Phys.\ A }{\bf 19}, 267-277.

\noindent
G\"ahler, F.\ (1986). {\it J.\ Physique, }{\bf 47}, C3-115--C3-122.

\noindent
G\"ahler, F.\ (1988). Crystallography of Dodecagonal Quasicrystals,
in {\it Quasicrystalline Materials}, C.\ Janot and J.M.\ Dubois (eds.),
World Scientific (Singapore).

\noindent
Gardner, M.\ (1977). {\it Sci.\ Am.\ }{\bf 236}, 110-121.

\noindent
Grunbaum, B.\ \& Shepard, G.\ C.\ (1987). {\it Tilings and Patterns, }
Freeman (San Francisco).

\noindent
Henley, C.\ L.\ (1986). {\it Phys.\ Rev.\ B }{\bf 34}, 797-810.

\noindent
Ishimasa, T.\, Nissen, H.-U.\ \& Fukano, Y.\ (1985).
{\it Phys.\ Rev.\ Lett.\ }{\bf 55}, 511-513.

\noindent
Jaric, M.\ V.\ (1986). {\it Phys.\ Rev.\ B }{\bf 34}, 4685-4689.

\noindent
Kalugin, P.\ A.\, Kitaev, A.\ Yu.\, \& Levitov, L.\ S.\ (1985).
{\it Zh.\ Eksp.\ Theor.\ Fiz.\ Pis.\ Red.\ }{\bf 41}, 119-121 (English
translation: {\it JETP Lett.\ }{\bf 41}, 119-121.

\noindent
Korepin, V.\ E.\ (1986). {\it Zap.\ Nauch.\ Semin.\ LOMI }{\bf 155}, 116-135 (in Russian).

\noindent
Kramer, P.\ \& Neri, R.\ (1984). {\it Acta Cryst. A}{\bf 40}, 580-587.

\noindent
Les Houches Proceedings (1986). {\it Proceedings of the International
Workshop on Aperiodic Crystals, J.\ Physique Colloque C3.}

\noindent
Levine, D.\ \& Steinhardt, P.\ J.\ (1984). {\it Phys.\ Rev.\ Lett.\ }
{\bf 53}, 2477-2480.

\noindent
Mackay, A.\ L.\ (1981). {Sov.\ Phys.\ Crystallogr.\ }{\bf 26}, 517-522.

\noindent
Penrose, R.\ (1974). {\it Bull.\ Inst.\ Math.\ Appl.\ }{\bf 10}, 266-271.

\noindent
Penrose, R.\ (1979). {\it Math.\ Intell.\ }{\bf 2}, 32-37.

\noindent
Robinson, R.\ M.\ (1971). {\it Invent.\ Math.\ }{\bf 12}, 177-209.

\noindent
Shechtman, D.\, Blech, I., Gratias D., \& Cahn, J.~W.\ (1984).
{\it Phys.\ Rev.\ Lett.\ }{\bf 53}, 1951-1953.

\noindent
Socolar, J.\ E.\ S.\, Steinhardt, P.\ J.\ \& Levine, D.\ (1985).
{\it Phys.\ Rev.\ B}{\bf 32}, 5547-5550.

\noindent
Stampfli, P.\ (1986). {\it Helv.\ Phys.\ Acta }{\bf 59}, 1260-1263.

\noindent
Wang, K.\ (1965). {\it Sci.\ Am.\ }{\bf 216}, 98-105.

\noindent
Zia, R.\ K.\ P.\ \& Dallas, W.\ J.\ (1985). {\it J.\ Phys.\ A}{\bf 18},
L341-L345.

\end{document}